\documentclass{article}
\usepackage{spconf,amsmath,graphicx}
\usepackage{hyperref}
\usepackage{mathtools}
\usepackage{tcolorbox}
\usepackage{color}
\usepackage{theorem}
\usepackage{times,amsmath,epsfig}
\usepackage{amssymb}
\usepackage{subfigure}
\usepackage{cite}
\usepackage{cases}
\usepackage{siunitx}
\usepackage{algorithmic}
\usepackage{relsize}
\usepackage{algorithm}
\usepackage{appendix}
\usepackage{needspace}

\newcommand{\myparagraph}[1]{\needspace{1\baselineskip}\medskip\noindent {\it #1.}}

\usepackage{tikz}
\usetikzlibrary{shapes,arrows}
\input{mysymbol.sty}
\tikzstyle{phantom vertex} = [ ellipse, 
                               anchor = center, 
                               minimum height = 1*\unit, 
                               minimum width  = 1*\unit,
                               inner sep=0pt,
                               anchor=center]
\tikzstyle{red vertex}   = [black, fill = red!20,   phantom vertex, draw]
\tikzstyle{black vertex} = [black, fill = black!20, phantom vertex, draw]
\tikzstyle{blue vertex}  = [black, fill = blue!20,  phantom vertex, draw]
\tikzstyle{green vertex} = [black, fill = green!20,  phantom vertex, draw]
\tikzstyle{yellow vertex} = [black, fill = yellow!20,  phantom vertex, draw]
\tikzstyle{cyan vertex} = [black, fill = cyan!20,  phantom vertex, draw]
\tikzstyle{vertex}       = [draw, phantom vertex]

\tikzstyle{point} = [ellipse, inner sep=0pt, draw, fill=white, anchor = center,
                     minimum height = 0.05*\unit, minimum width  = 0.05*\unit]

\newtheorem{myproposition}{\bf Proposition}
\newtheorem{mycorollary}{\bf Corollary}[myproposition]
\newtheorem{remark}{\bf Remark}

\title{Stability Analysis of Unfolded WMMSE for Power Allocation}
\name{Arindam Chowdhury, Fernando Gama, and Santiago Segarra 
\thanks{Research was sponsored by the Army Research Office and was accomplished under Cooperative Agreement Number W911NF-19-2-0269. 
The views and conclusions contained in this document are those of the authors and should not be interpreted as representing the official policies, either expressed or implied, of the Army Research Office or the U.S. Government. 
The U.S. Government is authorized to reproduce and distribute reprints for Government purposes notwithstanding any copyright notation herein.
\newline
Emails:  \{ac131, fgama, segarra\}@rice.edu.}} 
\address{Electrical and Computer Engineering, Rice University, USA  \hspace{1cm}}
\begin{document}
\ninept
\renewcommand{\baselinestretch}{0.95}
\maketitle
\begin{abstract}
%
Power allocation is one of the fundamental problems in wireless networks and a wide variety of algorithms address this problem from different perspectives. 
A common element among these algorithms is that they rely on an estimation of the channel state, which may be inaccurate on account of hardware defects, noisy feedback systems, and environmental and adversarial disturbances. 
Therefore, it is essential that the output power allocation of these algorithms is stable with respect to input perturbations, to the extent that the variations in the output are bounded for bounded variations in the input. 
In this paper, we focus on UWMMSE -- a modern algorithm leveraging graph neural networks --, and illustrate its stability to additive input perturbations of bounded energy through both theoretical analysis and empirical validation.
\end{abstract}
\begin{keywords}
Wireless networks, power allocation, UWMMSE, stability
\end{keywords}
\section{Introduction}
\label{sec:intro}

Power control is among the most fundamental resource allocation problems in communication systems~\cite{shannon1948mathematical,cave2007essentials}. 
It plays a crucial role in ensuring smooth functioning of physical systems and devices in both wired and wireless communication networks. 
Optimal allocation of power is especially critical under interference scenarios in multi-user ad hoc networks as devices in such networks tend to have limited availability of power while the channel characteristics and user demands can change fast depending on deployment scenarios~\cite{yu2016fair}. 
This necessitates solving a system-utility optimization problem with respect to allocated power under imposed constraints~\cite{boche2011characterization}.   

In spite of a straightforward formulation, this optimization problem turns out to be NP-hard~\cite{luo_2008_dynamic, hong_2014_signal}. 
Several broad classes of solution models have been proposed for this problem. 
These include classical iterative approaches based on block-coordinate descent~\cite{shi2011iteratively} and data-driven~\cite{roddenberry_2019_hodgenet,kumar2021adaptive,zhao2021distributed} connectionist approaches that approximate a power allocation policy through suitable parametrizations~\cite{eisen2019learning,shen2019graph,sun2018learning}.
While the iterative approaches have better convergence guarantees, they are quite slow and computationally complex resulting in a major bottleneck, especially under fast-changing channel conditions~\cite{chowdhury2021unfolding}. 
This is remedied by feed-forward neural models~\cite{qin2019deep,sun2018learning} which, in turn, suffer from standard deficiencies of neural networks like the unavailability of convergence guarantees~\cite{chowdhury2021unfolding}. 
More recently, a novel class of hybrid methods has emerged, combining key elements from both classes to generate fast as well as effective power allocation algorithms~\cite{hu2020iterative,pellaco2020deep,chowdhury2021unfolding}. 

Although these broad categories of methods employ a wide range of distinct algorithms and architectures, they usually have very similar input/output characteristics. 
The standard input to these methods is a set of channel estimates and interference values among network transceivers, while the output is the necessary power allocation~\cite{eisen2020optimal,shi2011iteratively,chowdhury2021unfolding,shen2019graph}.
Our objective is to analyze the input-output stability of these solution methods. 
The necessity of this analysis stems from the real-world scenarios wherein it is almost never possible to perfectly estimate the channel coefficients given various system-level  perturbations, namely, hardware defects, noisy feedback systems, external disturbances, and even adversarial attacks~\cite{goldsmith2005wireless,tse2005fundamentals,kyritsi2001effect}. 
Yet, it is essential that the power allocation algorithms are stable insofar as the variations in their output are bounded for bounded variations in corresponding input. Needless to say, an analysis of this form is both a prerequisite for deployment in real-world scenarios and an essential factor for comparison of various methods --in addition to their speed and effectiveness. 

A significant impediment to this analysis lies in the intractability and computational impracticability of generating a model for the physical disturbances in the system~\cite{kyritsi2001effect,yoo2006capacity}. As a workaround, we define our perturbation as a random variable, with a tractable parametric distribution, that can be applied to any power allocation algorithm.  In this work, we choose to focus on a particular hybrid architecture, the Unfolded WMMSE (UWMMSE)~\cite{chowdhury2021unfolding} and analyze its inference behaviour with respect to perturbations in its input.
UWMMSE has proven to be an effective algorithm for power-allocation, achieving near-optimal performance on both SISO and MIMO~\cite{chowdhury2021ml} test-cases. 
Importantly, it lies at the intersection of iterative and connectionist methods and thus an analysis of its operational structure provides intuition for both classes.

\medskip\noindent\textbf{Contribution.} 
The contributions of this paper are twofold:

i) We provide a theoretical stability result for UWMMSE under certain reasonable assumptions and verify it through simulations.

ii) We empirically illustrate the superiority of UWMMSE by comparisons with state-of-the-art alternatives, in terms of corresponding variations in outputs with respect to identical perturbations in inputs.  

\section{System model and problem formulation}\label{S:Modeling}
We limit the scope of our analysis to power allocation methods for SISO wireless interference networks. 
To that end, a single-hop ad hoc interference network with $M$ distinct single-antenna transceiver pairs is chosen to be our analytical and experimental testbed. 
The transmitters in the network are denoted by $i$ and the $i$-th transmitter can communicate only with its associated receiver, denoted by $r(i)$ for all $i \in \{1, \ldots, M\}$.  
The nodes in this network follow a linear transmission model of the form, 
\begin{equation}\label{E:trans_model}
     y_i = h_{ii}x_i + \sum_{\substack{j=1 \, | \,  j\neq i}}^M h_{ij}x_j + n_i,
\end{equation}
where $x_i$ and $y_i$ denote transmitted and received signals respectively at transceiver $i$, $h_{ii} \in \mathbb{R}$ is the channel between the $i$-th transceiver pair, $h_{ij} \in \mathbb{R}$ for $i \neq j$ represents the interference between transmitter $j$ and receiver $r(i)$, and $n_i \sim \ccalN(0, \sigma^2)$    
represents the additive channel noise. 
The instantaneous channel states are stored in a channel-state matrix $\bbH \in \reals^{M \times M}$ where $[\bbH]_{ij} = h_{ij}$.
Given the inherent topological structure of an ad hoc interference network, $\bbH$ can be interpreted as the weighted adjacency matrix of a directed graph with $M$ nodes.

In this setting, the instantaneous data rate $c_i$ at receiver $r(i)$ is given by Shannon's capacity theorem where the instantaneous power allocated to transmitter $i$ is given by $p_i \ge 0$ and $\bbp = [p_1, \ldots, p_M]^\top$.
The optimal power allocation is usually obtained by optimizing a network performance utility function within a power constraint. For example, the sum-rate maximization problem is defined as follows: 
\begin{align}\label{E:optimization_problem}
&\hspace{2.5cm} \max_{\bbp} \,\, \sum_{i=1}^M c_i(\bbp, \bbH) \qquad   \\ \
& \text{s.t.} \,\,\,\,\, c_i(\bbp, \bbH) = \log_2 \left( 1 + \frac{ h_{ii}^2 p_i}{\sigma^2 + \sum_{j\neq i} h_{ij}^2 p_j} \right), \nonumber\\ 
& \,\,\,\,\,\,\,\,\,\, 0 \leq p_i \leq p_{\max}, \,\, \text{for all} \,\, i, \nonumber
\end{align}
where $p_{\max}$ denotes the maximum available power.

While WMMSE, a classical approach to solve this problem, involves block-coordinate-descent based iterative updates~\cite{shi2011iteratively}, several connectionist and hybrid approaches have been proposed that learn a mapping from the channel-state matrix $\bbH$ to instantaneous power-allocation $\bbp$ directly from data. 
Although these methods differ in structure, the standard input to all the methods is $\bbH$ and their output tends to be either a deterministic allocation $\bbp$ or the parameters of a power allocation policy from which $\bbp$ can be sampled. 
In essence, we can define a generic feed-forward power allocation model as $\bbp = \bbPhi(\bbH; \bbTheta)$ wherein $\bbPhi$ is any user-chosen solution method and $\bbTheta$ is a set of trainable parameters, defined only for connectionist approaches. 
Note here that $\bbTheta$, if defined, would be a trained set of weights, meaning that we are only concerned with the inference-time behaviour of the trainable models under the assumptions that the weights have already been tuned for the power allocation task and that the training and inference data are independently sampled from identical (unobserved) distributions. 

Given a $\bbPhi$ and two channel matrices $\bbH, \tbH$, our objective in this work is to establish an elementwise upper bound on the variation of its output under a small bounded variation in its input. Namely,
\begin{equation} \label{eqn:objective}
    |[\bbPhi(\bbH;\bbTheta) - \bbPhi(\tbH;\bbTheta)]_{i}| \lesssim \eps
\end{equation}
where $\eps$ is a small number that represents the dissimilarity between the channels $\bbH$ and $\tbH$. In particular, we focus on $\bbPhi$ given by the Unfolded WMMSE (UWMMSE)~\cite{chowdhury2021unfolding}.

\section{UWMMSE architecture}
UWMMSE has a hybrid, connectionist architecture that modifies the iterative updates of WMMSE in two ways. First, the number of iterations is truncated to reduce computations and, second, the update trajectories of variables are accelerated for faster convergence. This results in a faster and more efficient algorithm that solves the following reformulated optimization problem: 
\begin{align}\label{E:problem_reformulation}
& \hspace{1.2cm} \min_{\mathbf{w,u,v}} \sum_{i=1}^M (w_i e_i - \log w_i )\\
& \text{s.t.} \,\,\,\,\, e_i = (1-u_i h_{ii} v_i)^2 + \sigma^2 u_i^2 + \sum_{i \neq j} u_i^2 h_{ij}^2 v_j^2, \nonumber\\ 
& \,\,\,\,\,\,\,\,\,\, v^2_i \leq p_{\max} \quad \text{for all } i,\nonumber
\end{align}
where $e_i$ is the mean-square error of the signal at node $i$, assuming that the transmitted signal is independent of channel noise. 
In this reformulation, $\bbu$ and $\bbv$ represent receiver and transmitter side characteristics, respectively, while $\bbw$ is a combining weight that represents the quality of transmission between a transceiver pair. 

It can be shown~\cite[Thm. 3]{shi2011iteratively} that an optimal solution $\{\bbw^*, \bbu^*, \bbv^*\}$ of~\eqref{E:problem_reformulation} generates an optimal solution $\bbp^*$ of~\eqref{E:optimization_problem} as $\bbp^* = (\bbv^*)^2$, with elementwise squares. 
Since~\eqref{E:problem_reformulation} is convex in each variable with the other two fixed, it can be solved using iterative block-coordinate descent, thus providing closed-form update rules for each variable. 

Unfolding this algorithm~\cite{liu2019deep,monga2019algorithm,farsad2020data} involves augmenting one or more of these variable updates with trainable parameters and fixing the number of iterations to form a cascade of hybrid layers that combine closed-form iterative structures with neural architectures. 
UWMMSE allocates power as a function of the channel state matrix $\bbp = \bbPhi(\bbH; \bbTheta)$ through an unfolded architecture $\bbPhi$ with trainable weights $\bbTheta$. 
More precisely, with an initial $\bbv^{(0)} = \sqrt{p_{\max}} \, \mathbf{1}$, each layer $k = 1, \ldots, K$ is defined as
\begin{align}
\bba^{(k)} &= \Psi(\bbH; \bbtheta_a^{(k)}),  \qquad \bbb^{(k)}  = \Psi(\bbH;\bbtheta_b^{(k)}), \hspace{-10mm} \label{E:unfold_1}\\
u^{(k)}_i &= \frac{h_{ii}v^{(k-1)}_i}{\sigma^2 + \sum_j h_{ij}^2 {v^{(k-1)}_j} v^{(k-1)}_j}, \,\, &&\text{for all } i, \label{E:unfold_2}\\
w^{(k)}_i &= \frac{a_i^{(k)}}{1 - u^{(k)}_i h_{ii} v^{(k-1)}_i} + b_i^{(k)}, &&\text{for all } i, \label{E:unfold_3}\\
v^{(k)}_i &= \alpha \left( \frac{ u^{(k)}_i h_{ii} w^{(k)}_i}{\sum_j  h_{ji}^2 u^{(k)}_j u^{(k)}_j w^{(k)}_j}\right), &&\text{for all } i,\label{E:unfold_4}
\end{align}
%
%
and the final power allocation is obtained as $\bbp = \bbPhi(\bbH; \bbTheta) = (\bbv^{(K)})^2$, with elementwise squares.
The parameters $\bba^{(k)}$ and $\bbb^{(k)}$ are learned using a graph neural network (GNN) architecture~\cite{gama2020gnns, ruiz2021gnns} and are used to augment the update structure of $\bbw$ to drive the algorithm towards faster convergence. 
A more detailed convergence analysis of this method can be found in~\cite[Thm. 1]{chowdhury2021unfolding}.     
The non-linearity $\alpha(z) \coloneqq [z]_0^{\sqrt{p_{\max}}}$ in~\eqref{E:unfold_4} restricts the values of $v_i^{(k)} \in [0, \sqrt{p_{\max}}]$ by saturating the raw output at the extreme values thereby satisfying the power constraint in~\eqref{E:problem_reformulation}.
The trainable functions $\Psi$ in~\eqref{E:unfold_1} are modelled as graph convolutional networks (GCNs)~\cite{kipf2016semi} parametrized using $\bbtheta_a^{(k)}$ and $\bbtheta_b^{(k)}$ for all $K$ layers.

\section{Stability of UWMMSE}\label{S:uwmmse}

Let $\bbH, \tbH \in \reals^{M \times M}$ be two channel matrices. 
We consider $\bbH$ to be the \emph{true} channel matrix of our system while $\tbH$ is the measured channel matrix. 
Note that $\tbH$ may differ from  $\bbH$ due to various system-level perturbations. 
In what follows, due to the intractability of accurately modeling these perturbations, we consider $\tbH$ to be a noisy version of $\bbH$. 
More specifically, let $\tbE$ be a random matrix such that $[\tbE]_{ij} \sim \ccalN(0,\tau^{2}) \ s.t\ \tau \to 0$, from which we construct a matrix $\bbE \in \reals^{M \times M}$ such that $[\bbE]_{ij} = [\tbE]_{ij}$ if $|[\tbE]_{ij}| \leq \eps$ and $[\bbE]_{ij} = \pm \eps$ otherwise. 
Then, we define $\tbH = \bbH + \bbE$ as the channel \emph{perturbation}. 
The objective is to show that $|[\bbPhi(\bbH;\bbTheta) - \bbPhi(\tbH;\bbTheta)]_{i}| \leq C \eps$ for some $C<\infty$ [cf. \eqref{eqn:objective}]. 
To do this, we bound the UWMMSE output $v_i^{(k)}$ for all $i$ at each iteration $k$, as stated in the following result.

\begin{myproposition}\label{P:mainRes}
Let $\bbPhi$ be an UWMMSE architecture~\eqref{E:unfold_1}-\eqref{E:unfold_4} with $K$ layers. 
Also, let $\bbH \in \mathbb{R}^{N \times N}$ and $\tbH = \bbH + \bbE$ be two channel matrices with $[\bbE]_{ij} \leq \eps$ for all transceiver pairs $i$ and $r(j)$.  Then, for every node $i$ at any arbitrary layer $k$, it holds that
    \begin{equation}\label{eq:lipsc}
    |v_i^{(k)} - \tilde{v}_i^{(k)}| 
    \leq \frac{1}{\left| D_{v_i}^{(k)}\right|} \Bigg(\mathcal{E}^{(k)}_{1_i} + \frac{\left| N_{\tilde{v}_i}^{(k)}\right|}{\left| D_{\tilde{v}_i}^{(k)}\right|}\mathcal{E}^{(k)}_{2_i}\Bigg) \\
    \end{equation}
where, $D_{v_i}^{(k)} = \sum_j  h_{ji}^2 (u^{(k)}_j)^2 w^{(k)}_j$, $ N_{\tilde{v}_i}^{(k)} = \tilde{u}^{(k)}_i \tilde{h}_{ii} \tilde{w}^{(k)}_i$, and $ D_{\tilde{v}_i}^{(k)} = \sum_j  \tilde{h}_{ji}^2 (\tilde{u}^{(k)}_j)^2 \tilde{w}^{(k)}_j$. 
Further, under certain system-level simplifying assumptions (to be specified in Remark~\ref{R:assumptions}), $\mathcal{E}^{(k)}_{1_i}$ and $\mathcal{E}^{(k)}_{2_i}$ can be obtained as,
\begin{align*}
\mathcal{E}^{(k)}_{1_i} &= \left| N_{v_i}^{(k)} - N_{\tilde{v}_i}^{(k)}\right| = \frac{h_{ii}^2}{\left|D_{u_i}^{(k)}\right|} \bigg( \left| v_i^{(k-1)} - \tilde{v}_i^{(k-1)}\right| \\ &\hspace{0.5cm} + \frac{h_{ii}^2}{\left| D_{u_i}^{(k)}\right|} \left|{a_i^{(k)} \{v_i^{(k-1)}\}^3} - \tilde{a}_i^{(k)} \{\tilde{v}_i^{(k-1)}\}^3\right| \bigg) \\ &\hspace{0.5cm} + \ell(\eps, h_{ii}, v_{i}^{(k)}, D_{u_i}^{(k)}, \tilde{a}_{i}^{(k)}) + \mathcal{O}(\eps^2 +\eps^3 + \dots)
\end{align*}
\begin{align*}
\mathcal{E}^{(k)}_{2_i} &= \left| D_{v_i}^{(k)} - D_{\tilde{v}_i}^{(k)}\right| = \mathlarger{\mathlarger{\sum}}_j \left|h_{jj}\right| \Bigg[ \frac{h_{ji}^2 }{\left|D_{u_j}^{(k)}\right|} \Bigg(\left| v_j^{(k-1)} - \tilde{v}_j^{(k-1)}\right| \\ &\hspace{0.5cm} + \frac{h_{jj}^2}{\left|D_{u_j}^{(k)}\right|} \left| a_j^{(k)}\{v_j^{(k-1)}\}^3 - \tilde{a}_j^{(k)}\{\tilde{v}_j^{(k-1)}\}^3\right|\Bigg) \\ &\hspace{0.5cm} + \ell(\eps, h_{jj}, h_{ji} v_{j}^{(k-1)}, D_{u_j}^{(k)}, \tilde{a}_{j}^{(k)}) + \mathcal{O}(\eps^2+\eps^3+\dots)
\end{align*}
where $\ell(\eps,\cdot)$ are linear terms on $\eps$ whose exact forms are omitted for simplicity.
\end{myproposition}
A proof of the proposition is provided in the Appendix. The above result provides an upper bound on the per-layer variation of UWMMSE's output. 
This upper bound is composed of multiple factors, including strengths of perturbed channels, given by $N_{\tilde{v}_i}^{(k)}$, interference components, given by $D_{v_i}^{(k)}, D_{\tilde{v}_i}^{(k)}$ and also the layer-wise errors $\{\mathcal{E}^{(k)}_{1_i}, \mathcal{E}^{(k)}_{2_i}\}$.
In addition to being functions of the input perturbation $\eps$, $\mathcal{E}^{(k)}_{1_i}$ and $\mathcal{E}^{(k)}_{2_i}$ depend directly on errors accumulating from the previous layers. 
These layer-wise errors essentially quantify the absolute change in channel strengths and interference components -- which directly impact the power allocation process -- due to input perturbation and thus play a key role in defining an upper bound, albeit loose, on the change in $v^{(k)}$ at any $k$.    

A special case arises for number of layers $K = 1$, where inputs $v_i^{(0)} = \tilde{v}_i^{(0)} = \sqrt{p_{\max}}$. This is formalized in the following corollary.
\begin{mycorollary}\label{C:cor1}
    If $K = 1$, then for all $i$, $\mathcal{E}^{(k)}_{1_i}$ and $\mathcal{E}^{(k)}_{2_i}$ reduce to
    \begin{align*}
        \mathcal{E}^{(k)}_{1_i} &= \frac{h_{ii}^4 \{\sqrt{p_{\max}}\}^3}{\{D_{u_i}^{(k)}\}^2} \bigg( \left|a_i^{(k)} - \tilde{a}_i^{(k)} \right| \bigg) \\ &+ \ell(\eps, h_{ii}, D_{u_j}^{(k)}, \sqrt{p_{\max}}, \tilde{a}_{i}) + \mathcal{O}(\eps^2+\eps^3+\dots)
    \end{align*}
    \begin{align*}
        \mathcal{E}^{(k)}_{2_i} &= \mathlarger{\mathlarger{\sum}}_j \frac{\left|h_{jj}\right| (h_{ji} h_{jj})^2 \{\sqrt{p_{\max}}\}^3}{\left|D_{u_j}^{(k)}\right|^2} \Bigg( \left| a_j^{(k)} - \tilde{a}_j^{(k)}\right| \Bigg) \\ &+ \ell(\eps, h_{jj}, h_{ji}, D_{u_j}^{(k)}, \sqrt{p_{\max}}, \tilde{a}_{j}^{(k)}) + \mathcal{O}(\eps^2+\eps^3+\dots)
    \end{align*}
\end{mycorollary}
Clearly, the expressions for the propagation errors reduce to more amenable forms when $K=1$ but become progressively involved as $K$ increases. 
Nevertheless, the simplified expressions are of definite importance due to their explicit dependencies on variations in the learnable parameters $a^{(k)}$. 
It can be shown~\cite[Thm. 4]{gama2020stability} that $\left|a^{(k)} - \tilde{a}^{(k)} \right| = \left|\Psi(\bbH; \bbtheta_a^{(k)}) - \Psi(\tbH; \bbtheta_a^{(k)})\right| \leq B_\eps$ for all $k$ under the conditions that $\Psi$ is a GNN architecture and $\| \bbH - \tbH \|_{2} \leq \eps$ which are both satisfied in this case. 
Therefore, for $K=1$, the upper bounds for the propagation errors depend majorly on the characteristics of the neural network architecture $\Psi$.
Intuitively, the error in the first layer can be primarily attributed to the learnable components since the error due to unfolding can only accumulate with an increase in the number of layers.  
A similar result also holds for $b^{(k)} = \Psi(\bbH; \bbtheta_b^{(k)})$. 
However, under the simplifying assumptions used in Proposition~\ref{P:mainRes} (and detailed below in Remark~\ref{R:assumptions}), the effect of parameter $b^{(k)}$ in the propagation errors can be neglected.        

\begin{remark}[Simplifying assumptions]\label{R:assumptions}
\normalfont 
While the above results offer useful insights into the error propagation mechanism across the layers of UWMMSE, it is also important to analyze the key simplifying assumptions under which these results are valid. 
Firstly, we assume that the system operates in the high SNR regime which renders the noise variance $\sigma^2 \approx 0$ in~\eqref{E:unfold_2}. 
This is the preferred analytical setup for power allocation methods on account of the more challenging interference-management scenario that it offers. 
Secondly, we reformulate~\eqref{E:unfold_3} to
\begin{align*}
    w^{(k)}_i &= \big( a_i^{(k)} + b_i^{(k)} \big)  + a_i^{(k)}u^{(k)}_i h_{ii} v^{(k-1)}_i , &&\text{for all } i.
    \intertext{Since it can be empirically shown that $u^{(k)}_i h_{ii} v^{(k-1)}_i \leq 1$, the above is achieved using the Taylor's expansion of $(1 - u^{(k)}_i h_{ii} v^{(k-1)}_i)^{-1}$. Further, by enforcing $a_i^{(k)}+b_i^{(k)} = 1$ as a design choice, the update step is reduced to}
    w^{(k)}_i &= 1  + a_i^{(k)}u^{(k)}_i h_{ii} v^{(k-1)}_i , &&\text{for all } i.
\end{align*}    
Finally, we assume that $D_{u_i}^{(k)} \approx D_{\tilde{u}_i}^{(k)}$. While this may appear to be an unreasonable assumption at first, careful consideration of $D_{\tilde{u}_i}^{(k)} = \sum_j  \tilde{h}_{ij}^2 (\tilde{v}^{(k-1)}_j)^2$ reveals that $D_{\tilde{u}_i}^{(k)} = \sum_j  h_{ij}^2 (\tilde{v}^{(k-1)}_j)^2 + \mathcal{O}(\eps + \eps^2)$, and this last term can be neglected for small $\eps$. 
And, for constrained power allocations, wherein the input $v^{(k-1)}$ to each step is bounded by $\alpha(\cdot)$ which is usually a tight bound as $p_{\max}$ tends to be small in real-world scenarios, the difference between $v^{(k-1)}$ and $\tilde{v}^{(k-1)}$ is not too big and our assumption is reasonably justified. 
\end{remark}

\section{Numerical experiments}\label{S:num_exp}
\begin{figure*}[!htbp]
	\centering
	\subfigure[]{
			\centering
			\includegraphics[width=0.31\textwidth]{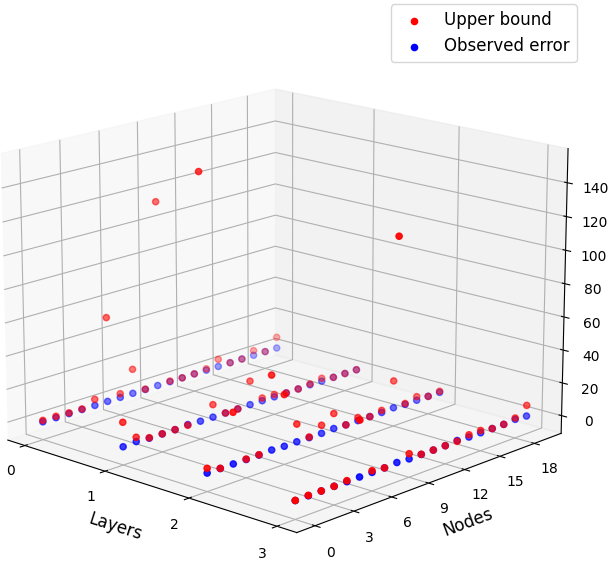}
			\label{Fig:val}
		}	
	\subfigure[]{
			\centering
			\includegraphics[width=0.31\textwidth]{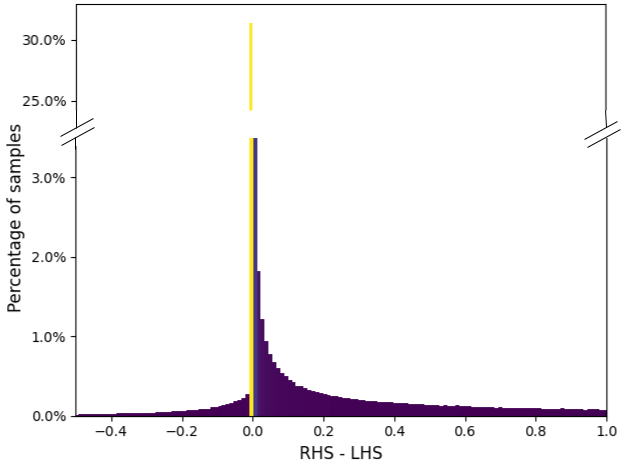}
			\label{Fig:hist}
		}	
	\subfigure[]{
			\centering
			\includegraphics[width=0.31\textwidth,height=0.235\textwidth]{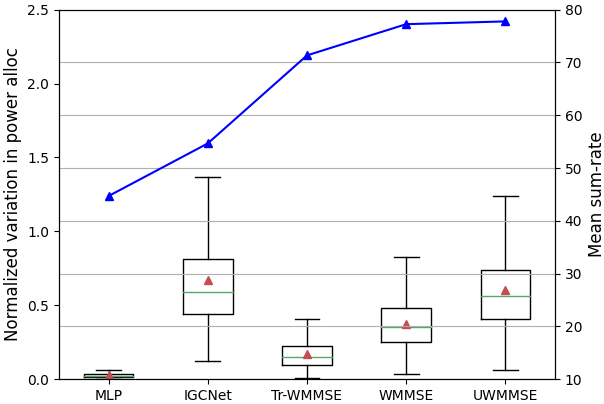}
			\label{Fig:com}
		}
		\caption{\small {Stability and performance analysis of UWMMSE and baselines. 
		(a)~3-D plot showing upper bound (RHS) and observed errors (LHS) of~\eqref{eq:lipsc} for a random $\bbH,\tbH$ pair with $20$ nodes and $4$ unrolling layers of UWMMSE.
		(b)~Truncated histogram of difference of upper bounds and observed errors for $800$k test nodes using UWMMSE. 
		(c)~Box plots of normalized variation in output and plot of corresponding mean sum-rates (blue curve) achieved by MLP, IGCNet, Tr-WMMSE, WMMSE and UWMMSE.}
		}
		\label{Fig:comparison}
\end{figure*}

We choose a Geometric-Rayleigh fading channel model for our experiments. 
Identical to~\cite{chowdhury2021efficient,chowdhury2021ml}, transmitters and receivers are dropped uniformly at random in a bounded geometric space and the corresponding channel coefficients are proportional to the inverse of the distance between pairs of transceivers and a fading component sampled from a Rayleigh distribution.  
For training, we follow the operating conditions as described in~\cite{chowdhury2021efficient}.

For evaluation\footnote{Code to replicate the numerical experiments can be found at \href{https://github.com/ArCho48/stability-UWMMSE.git}{https://github.com/ArCho48/stability-uwmmse.git}.}, we generate a test set consisting of $10,\!000$ i.i.d channel realizations.
The elementwise perturbations are sampled from a standard normal distribution $\mathcal{N}(0,1)$ and clipped at an absolute maximum of $0.005$. 
We now provide an empirical validation of Proposition~\ref{P:mainRes}, and then present a comparison of UWMMSE with established baselines in terms of variations in output with respect to identically generated inputs.   


\myparagraph{\textbf{Empirical validation}} For validating Proposition~\ref{P:mainRes}, we choose to compare the theoretical upper bound (excluding the $\ell(\cdot)$ and the $\mathcal{O}(\eps^2+\dots)$ terms) for each node in the test set with the observed error. 
Figure~\ref{Fig:val} shows a 3-D plot for a randomly chosen $\bbH$ from the test set, perturbed with $\left|\eps\right| \leq 0.005$. 
Clearly, the upper bound is satisfied by the majority of the $20$ nodes in the network at each of the $4$ unfolded layers. 
While the upper bound appears quite loose for a small set of nodes, it shows a tight fit for a sizeable section of nodes, given the resolution. 
To verify this further, we plot a histogram of differences between RHS and LHS of~\eqref{eq:lipsc} for $800$k nodes ($10,\!000$ realizations of $\bbH$ with $20$ nodes at each of the $4$ unrolled layers) in Figure~\ref{Fig:hist}. 
Here, we observe that for about $66\%$ of the nodes the upper bound is satisfied with a margin close to $0.0$ indicating tight fit, while about $30\%$ exhibit a loose fit with margin greater than $2.0$. 
More importantly, the number of violations (attributable to ignoring the higher order terms) is just about about $4\%$.
This shows the empirical validity of the established bounds even under simplifying assumptions.


\myparagraph{\textbf{Performance comparison}} Here, we compare multiple power allocation algorithms in terms of stability and performance. 
For evaluating stability, we choose the normalized variation in their final output as the metric. 
Formally, this is given by $\frac{\lVert\bbPhi(\tbH;\cdot) - \bbPhi(\bbH;\cdot)\rVert_2}{\lvert\bbPhi(\bbH;\cdot)\rVert_2}$.
Performance comparison is given in terms of sum-rate utility~\cite{shi2011iteratively}.
For this comparison, we choose {MLP}~\cite{sun2018learning}, {IGCNet}~\cite{shen2019graph}, WMMSE~\cite{shi2011iteratively} and its truncated version (Tr-WMMSE), which has the same number of iterations as the number of layers in UWMMSE~\cite{chowdhury2021unfolding}. 
Although this is not an exhaustive list of available power allocation algorithms, these methods cover the entire spectrum of connectionist and classical methods and are also a subset of the common baselines that UWMMSE was originally compared against~\cite{chowdhury2021efficient}. 
We note that this study does not include REGNN~\cite{eisen2020optimal} (a common baseline) since it uses a probabilistic policy for power allocation and our analysis in its current form is not suited for one-to-many mappings between unperturbed inputs and outputs.

As shown in Figure~\ref{Fig:com}, the mean normalized output variation of the hybrid UWMMSE lies somewhere in the middle of IGCNet, which is a purely connectionist method and the classical WMMSE.
This indicates a trade-off such that IGCNet, being purely data-driven, is slightly more prone to high variance w.r.t changes in input while UWMMSE having a strong structural bias inherent to the classical update structure, is a little more stable. 
At the same time, MLP, in spite of being purely data-driven, shows the least variation ($\approx0$). 
But its poor sum-rate performance (blue curve) proves that it is inadequate to capture the complexities of the power allocation process. These observations indicate that connectionist methods often find it challenging to balance the bias-variance trade-off required for accurate learning. 
Tr-WMMSE exhibits maximum stability among the rest and stability decreases as more iterations are added to reach the full WMMSE, thus illustrating the per-layer error propagation. 
UWMMSE, in spite of having as many layers as Tr-WMMSE, accumulates more error due to its embedded GCN~\cite{kipf2016semi} based learning modules.  
However, it achieves a mean sum-rate performance that is significantly better than all other methods and even marginally surpasses that of near-optimal WMMSE for this test-set, thus proving the superiority of hybrid algorithms.        

\section{Conclusions}\label{S:Conclusions}

We studied the stability properties of the UWMMSE power allocation algorithm. 
The main contribution of this work is a theoretical analysis that aims to establish an upper bound on the variation in per-layer output of UWMMSE w.r.t bounded variation in its input. 
This analysis provides useful insights into the operational structure of a superior class of hybrid methods. 
The theoretical results are also emphasized using supporting empirical evidence. 
We further demonstrate the effectiveness of UWMMSE by an experimental comparison of the variation in its final output with that of other state-of-the-art methods. 
Current work involves extending this analysis to more realistic perturbation models and adversarial scenarios. 



\bibliographystyle{IEEEbib}
\bibliography{strings,refs}

\begin{thebibliography}{10}

\bibitem{shannon1948mathematical}
Claude~E Shannon,
\newblock ``A mathematical theory of communication,''
\newblock {\em The Bell system technical journal}, vol. 27, no. 3, pp.
  379--423, 1948.

\bibitem{cave2007essentials}
Martin Cave, Christopher Doyle, and William Webb,
\newblock {\em Essentials of Modern Spectrum Management},
\newblock Cambridge University Press Cambridge, 2007.

\bibitem{yu2016fair}
Qin Yu, Yizhe Zhao, Lanxin Zhang, Kun Yang, and Supeng Leng,
\newblock ``A fair resource allocation algorithm for data and energy integrated
  communication networks,''
\newblock {\em Energy}, vol. 1, pp. U2, 2016.

\bibitem{boche2011characterization}
Holger Boche, Siddharth Naik, and Tansu Alpcan,
\newblock ``Characterization of convex and concave resource allocation problems
  in interference coupled wireless systems,''
\newblock {\em IEEE Trans. Signal Process.}, vol. 59, no. 5, pp. 2382--2394,
  2011.

\bibitem{luo_2008_dynamic}
Zhi-Quan Luo and Shuzhong Zhang,
\newblock ``Dynamic spectrum management: Complexity and duality,''
\newblock {\em IEEE J. Sel. Topics Signal Process.}, vol. 2, no. 1, pp. 57--73,
  2008.

\bibitem{hong_2014_signal}
Mingyi Hong and Zhi-Quan Luo,
\newblock ``Chapter 8 - {Signal Processing and Optimal Resource Allocation for
  the Interference Channel},''
\newblock in {\em Academic Press Library in Signal Processing: Volume 2},
  Nicholas~D. Sidiropoulos, Fulvio Gini, Rama Chellappa, and Sergios
  Theodoridis, Eds., vol.~2 of {\em Academic Press Library in Signal
  Processing}, pp. 409 -- 469. Elsevier, 2014.

\bibitem{shi2011iteratively}
Qingjiang Shi, Meisam Razaviyayn, Zhi-Quan Luo, and Chen He,
\newblock ``An iteratively weighted {MMSE} approach to distributed sum-utility
  maximization for a {MIMO} interfering broadcast channel,''
\newblock {\em IEEE Trans. Signal Process.}, vol. 59, no. 9, pp. 4331--4340,
  2011.

\bibitem{roddenberry_2019_hodgenet}
T.~Mitchell Roddenberry and Santiago Segarra,
\newblock ``{HodgeNet}: Graph neural networks for edge data,''
\newblock in {\em Asilomar Conf. Signals, Systems, and Computers}, 2019, pp.
  220--224.

\bibitem{kumar2021adaptive}
Abhishek Kumar, Gunjan Verma, Chirag Rao, Ananthram Swami, and Santiago
  Segarra,
\newblock ``Adaptive contention window design using deep q-learning,''
\newblock in {\em IEEE Intl. Conf. Acoust., Speech and Signal Process.
  (ICASSP)}, 2021, pp. 4950--4954.

\bibitem{zhao2021distributed}
Zhongyuan Zhao, Gunjan Verma, Chirag Rao, Ananthram Swami, and Santiago
  Segarra,
\newblock ``Distributed scheduling using graph neural networks,''
\newblock in {\em IEEE Intl. Conf. Acoust., Speech and Signal Process.
  (ICASSP)}, 2021, pp. 4720--4724.

\bibitem{eisen2019learning}
Mark Eisen, Clark Zhang, Luiz~FO Chamon, Daniel~D Lee, and Alejandro Ribeiro,
\newblock ``Learning optimal resource allocations in wireless systems,''
\newblock {\em IEEE Trans. Signal Process.}, vol. 67, no. 10, pp. 2775--2790,
  2019.

\bibitem{shen2019graph}
Yifei Shen, Yuanming Shi, Jun Zhang, and Khaled~B Letaief,
\newblock ``A graph neural network approach for scalable wireless power
  control,''
\newblock {\em arXiv preprint arXiv:1907.08487}, 2019.

\bibitem{sun2018learning}
Haoran Sun, Xiangyi Chen, Qingjiang Shi, Mingyi Hong, Xiao Fu, and Nicholas~D
  Sidiropoulos,
\newblock ``Learning to optimize: Training deep neural networks for
  interference management,''
\newblock {\em IEEE Trans. Signal Process.}, vol. 66, no. 20, pp. 5438--5453,
  2018.

\bibitem{chowdhury2021unfolding}
Arindam Chowdhury, Gunjan Verma, Chirag Rao, Ananthram Swami, and Santiago
  Segarra,
\newblock ``Unfolding wmmse using graph neural networks for efficient power
  allocation,''
\newblock {\em IEEE Trans. Wireless Commun.}, vol. 20, no. 9, pp. 6004--6017,
  2021.

\bibitem{qin2019deep}
Zhijin Qin, Hao Ye, Geoffrey~Ye Li, and Biing-Hwang~Fred Juang,
\newblock ``Deep learning in physical layer communications,''
\newblock {\em IEEE Wirel. Commun.}, vol. 26, no. 2, pp. 93--99, 2019.

\bibitem{hu2020iterative}
Qiyu Hu, Yunlong Cai, Qingjiang Shi, Kaidi Xu, Guanding Yu, and Zhi Ding,
\newblock ``Iterative algorithm induced deep-unfolding neural networks:
  Precoding design for multiuser mimo systems,''
\newblock {\em IEEE Trans. Wireless Commun.}, vol. 20, no. 2, pp. 1394--1410,
  2020.

\bibitem{pellaco2020deep}
Lissy Pellaco, Mats Bengtsson, and Joakim Jald{\'e}n,
\newblock ``{Deep unfolding of the weighted MMSE beamforming algorithm},''
\newblock {\em arXiv preprint arXiv:2006.08448}, 2020.

\bibitem{eisen2020optimal}
Mark Eisen and Alejandro~R Ribeiro,
\newblock ``Optimal wireless resource allocation with random edge graph neural
  networks,''
\newblock {\em IEEE Trans. Signal Process.}, 2020.

\bibitem{goldsmith2005wireless}
Andrea Goldsmith,
\newblock {\em Wireless Communications},
\newblock Cambridge university press, 2005.

\bibitem{tse2005fundamentals}
David Tse and Pramod Viswanath,
\newblock {\em Fundamentals of Wireless Communication},
\newblock Cambridge university press, 2005.

\bibitem{kyritsi2001effect}
Persefoni Kyritsi, Reinaldo~A Valenzuela, and Donald~C Cox,
\newblock ``Effect of the channel estimation on the accuracy of the capacity
  estimation,''
\newblock in {\em IEEE Veh. Technol. Conf. (VTC)}, 2001, vol.~1, pp. 293--297.

\bibitem{yoo2006capacity}
Taesang Yoo and Andrea Goldsmith,
\newblock ``Capacity and power allocation for fading mimo channels with channel
  estimation error,''
\newblock {\em IEEE Trans. Inf. Theory}, vol. 52, no. 5, pp. 2203--2214, 2006.

\bibitem{chowdhury2021ml}
Arindam Chowdhury, Gunjan Verma, Chirag Rao, Ananthram Swami, and Santiago
  Segarra,
\newblock ``Ml-aided power allocation for tactical {MIMO},''
\newblock {\em arXiv preprint arXiv:2109.06992}, 2021.

\bibitem{liu2019deep}
Risheng Liu, Shichao Cheng, Long Ma, Xin Fan, and Zhongxuan Luo,
\newblock ``Deep proximal unrolling: Algorithmic framework, convergence
  analysis and applications,''
\newblock {\em IEEE Trans. Image Process.}, vol. 28, no. 10, pp. 5013--5026,
  2019.

\bibitem{monga2019algorithm}
Vishal Monga, Yuelong Li, and Yonina~C Eldar,
\newblock ``Algorithm unrolling: Interpretable, efficient deep learning for
  signal and image processing,''
\newblock {\em arXiv preprint arXiv:1912.10557}, 2019.

\bibitem{farsad2020data}
Nariman Farsad, Nir Shlezinger, Andrea~J Goldsmith, and Yonina~C Eldar,
\newblock ``Data-driven symbol detection via model-based machine learning,''
\newblock {\em arXiv preprint arXiv:2002.07806}, 2020.

\bibitem{gama2020gnns}
Fernando Gama, Elvin Isufi, Geert Leus, and Alejandro Ribeiro,
\newblock ``Graphs, convolutions, and neural networks: From graph filters to
  graph neural networks,''
\newblock {\em {IEEE} Signal Process. Mag.}, vol. 37, no. 6, pp. 128--138, Nov.
  2020.

\bibitem{ruiz2021gnns}
Luana Ruiz, Fernando Gama, and Alejandro Ribeiro,
\newblock ``Graph neural networks: Architectures, stability and
  transferability,''
\newblock {\em Proc. {IEEE}}, vol. 109, no. 5, pp. 660--682, May 2021.

\bibitem{kipf2016semi}
Thomas~N Kipf and Max Welling,
\newblock ``Semi-supervised classification with graph convolutional networks,''
\newblock in {\em Intl. Conf. Learn. Repres. (ICLR)}, 2017.

\bibitem{gama2020stability}
Fernando Gama, Joan Bruna, and Alejandro Ribeiro,
\newblock ``Stability properties of graph neural networks,''
\newblock {\em IEEE Trans. Signal Process.}, vol. 68, pp. 5680--5695, 2020.

\bibitem{chowdhury2021efficient}
Arindam Chowdhury, Gunjan Verma, Chirag Rao, Ananthram Swami, and Santiago
  Segarra,
\newblock ``Efficient power allocation using graph neural networks and deep
  algorithm unfolding,''
\newblock in {\em IEEE Intl. Conf. Acoust., Speech and Signal Process.
  (ICASSP)}, 2021, pp. 4725--4729.

\end{thebibliography}

 \onecolumn
 \section*{\centering Appendix}

 \textbf{Proof of Propostion 1:}\\ 

 \noindent Let $v_i^{(k)} = \bbPhi(\bbH;\bbTheta)$ and $\tilde{v}_i^{(k)} = \bbPhi(\tbH;\bbTheta)$. Then, from Eq.~\eqref{E:unfold_4},
 \begin{align*}
     |v_i^{(k)} - \tilde{v}_i^{(k)}| &= \left| \alpha \left( \frac{ u^{(k)}_i h_{ii} w^{(k)}_i}{\sum_j  h_{ji}^2 \{u^{(k)}_j\}^2 w^{(k)}_j}\right) - \alpha \left( \frac{ \tilde{u}^{(k)}_i \tilde{h}_{ii} \tilde{w}^{(k)}_i}{\sum_j  \tilde{h}_{ji}^2 \{\tilde{u}^{(k)}_j\}^2 \tilde{w}^{(k)}_j}\right) \right| \\
     &\leq  \left| \frac{ u^{(k)}_i h_{ii} w^{(k)}_i}{\sum_j  h_{ji}^2 \{u^{(k)}_j\}^2 w^{(k)}_j} - \frac{ \tilde{u}^{(k)}_i \tilde{h}_{ii} \tilde{w}^{(k)}_i}{\sum_j  \tilde{h}_{ji}^2 \{\tilde{u}^{(k)}_j\}^2 \tilde{w}^{(k)}_j} \right| \tag*{since $\alpha$, being a clipping non-linearity, is $1$-lipschitz continuous}\\
     &\leq  \left| \frac{N_{v_i}^{(k)}}{D_{v_i}^{(k)}} - \frac{N_{\tilde{v}_i}^{(k)}}{D_{\tilde{v}_i}^{(k)}} \right| \tag*{where $N$ and $D$ represent the numerators and the denominators from the previous step}\\
     &\leq  \Bigg\{ \frac{\left| N_{v_i}^{(k)} - N_{\tilde{v}_i}^{(k)}\right|}{\left| D_{v_i}^{(k)}\right|} - \frac{\left| N_{\tilde{v}_i}^{(k)}\right|\left|D_{v_i}^{(k)} - D_{\tilde{v}_i}^{(k)} \right|}{\left|D_{v_i}^{(k)}\right| \left|D_{\tilde{v}_i}^{(k)}\right|} \Bigg\} \\
     \intertext{Now solving one fraction at a time,}
     &  \frac{\left| N_{v_i}^{(k)} - N_{\tilde{v}_i}^{(k)}\right|}{\left| D_{v_i}^{(k)}\right|}\\
     &\leq  \frac{\left| u^{(k)}_i h_{ii} w^{(k)}_i - \tilde{u}^{(k)}_i \tilde{h}_{ii} \tilde{w}^{(k)}_i \right|}{\left| D_{v_i}^{(k)}\right|}\\ 
     &\leq  \frac{ \left| u^{(k)}_i h_{ii}\{a_i +b_i + a_i h_{ii} u_i^{(k)} v_i^{(k-1)}\} - \tilde{u}^{(k)}_i \tilde{h}_{ii}\{\tilde{a}_i +\tilde{b}_i + \tilde{a}_i \tilde{h}_{ii} \tilde{u}_i^{(k)} \tilde{v}_i^{(k-1)}\} \right|}{\left| D_{v_i}^{(k)}\right|} \tag*{since $u_i h_{ii} v_i < 1, w_i \approx a_i(1 + u_i h_{ii} v_i) + b_i$ by Taylor's expansion}\\
     &\leq  \frac{\left| u^{(k)}_i h_{ii}\{1 + a_i h_{ii} u_i^{(k)} v_i^{(k-1)}\} - \tilde{u}^{(k)}_i \tilde{h}_{ii}\{1 + \tilde{a}_i \tilde{h}_{ii} \tilde{u}_i^{(k)} \tilde{v}_i^{(k-1)}\} \right|}{\left| D_{v_i}^{(k)}\right|} \tag*{$a_i + b_i$ can be made $1$ by design}\\
     \intertext{For the low-noise regime that we consider, channel noise variance $\sigma^2 \approx 0$. Therefore, $u^{(k)}_i = \frac{h_{ii}v^{(k-1)}_i}{D_{u_i}^{(k)}}$ where $ D_{u_i}^{(k)} = {\sum_{j} h_{ij}^2 v^{(k-1)}_j v^{(k-1)}_j}$. Again, assuming $D_{u_i}^{(k)} = D_{\tilde{u}_i}^{(k)}$}
     &\leq  \frac{\left|h_{ii}^2 v_i^{(k-1)} - \tilde{h}_{ii}^2 \tilde{v}_i^{(k-1)}\right|}{\left| D_{u_i}^{(k)}\right| \left| D_{v_i}^{(k)}\right|} + \frac{\left|a_i^{(k)} h_{ii}^4 \{v_i^{(k-1)}\}^3 - \tilde{a}_i^{(k)} \tilde{h}_{ii}^4 \{\tilde{v}_i^{(k-1)}\}^3\right|}{\{D_{u_i}^{(k)}\}^2\left| D_{v_i}^{(k)}\right|}\\
     &\leq \frac{}{\left|D_{u_i}^{(k)}\right| \left| D_{v_i}^{(k)}\right|}  \left\{ \left|h_{ii}^2 v_i^{(k-1)} - \tilde{h}_{ii}^2 \tilde{v}_i^{(k-1)}\right| + \frac{1}{\left|D_{u_i}^{(k)}\right|} \left|{a_i^{(k)} h_{ii}^4 \{v_i^{(k-1)}\}^3} - \tilde{a}_i^{(k)} \tilde{h}_{ii}^4 \{\tilde{v}_i^{(k-1)}\}^3\right| \right\}\\
     &\leq \frac{}{\left|D_{u_i}^{(k)}\right| \left| D_{v_i}^{(k)}\right|} \left\{ h_{ii}^2 \left| v_i^{(k-1)} - \tilde{v}_i^{(k-1)}\right| + 2\left| h_{ii} \epsilon_{ii}v_i^{(k-1)}\right| + \frac{1}{\left|D_{u_i}^{(k)}\right|} \left|{a_i^{(k)} h_{ii}^4 \{v_i^{(k-1)}\}^3} - \tilde{a}_i^{(k)} \tilde{h}_{ii}^4 \{\tilde{v}_i^{(k-1)}\}^3\right| + \mathcal{O}(\epsilon^2)\right\} \tag*{since $\tilde{h}_{ii} = h_{ii} + \epsilon_{ii}$}\\
     &\leq \frac{ h_{ii}^2}{\left|D_{u_i}^{(k)}\right| \left| D_{v_i}^{(k)}\right|} \left\{ \left| v_i^{(k-1)} - \tilde{v}_i^{(k-1)}\right| + \frac{h_{ii}^2}{\left| D_{u_i}^{(k)}\right|} \left|{a_i^{(k)} \{v_i^{(k-1)}\}^3} - \tilde{a}_i^{(k)} \{\tilde{v}_i^{(k-1)}\}^3\right| \right\}\\ &\hspace{5.0cm} + \frac{2}{\left|D_{u_i}^{(k)}\right| \left| D_{v_i}^{(k)}\right|}\left\{\left| h_{ii} \epsilon_{ii}v_i^{(k-1)}\right| + \frac{2}{\left| D_{u_i}^{(k)}\right|}\left| h_{ii}^3 \epsilon_{ii}\tilde{a}_i^{(k)}\{v_i^{(k-1)}\}^3 \right| + \mathcal{O}(\epsilon^2+\epsilon^3+\dots)\right\} \tag*{since $\tilde{h}_{ii} = h_{ii} + \epsilon_{ii}$} \\
     \intertext{And,}
     &  \frac{\left| N_{\tilde{v}_i}^{(k)}\right|\left|D_{v_i}^{(k)} - D_{\tilde{v}_i}^{(k)} \right|}{\left|D_{v_i}^{(k)}\right| \left|D_{\tilde{v}_i}^{(k)}\right|}\\
     &\leq \frac{ \left| N_{\tilde{v}_i}^{(k)}\right|}{\left|D_{v_i}^{(k)}\right| \left|D_{\tilde{v}_i}^{(k)}\right|} \left|\sum_j h_{ji}^2 \{u^{(k)}_j\}^2 w^{(k)}_j  - \tilde{h}_{ji}^2 \{\tilde{u}^{(k)}_j\}^2 \tilde{w}^{(k)}_j \right|\\
     \intertext{Expanding as in the previous case with the assumption that $D_{u_i}^{(k)} = D_{\tilde{u}_i}^{(k)}$,}
     &\leq \frac{ \left| N_{\tilde{v}_i}^{(k)}\right|}{\left|D_{v_i}^{(k)}\right| \left|D_{\tilde{v}_i}^{(k)}\right|} \mathlarger{\mathlarger{\sum}}_j \Bigg[ \frac{h_{ji}^2 \left|h_{jj}\right|}{\left|D_{u_j}^{(k)}\right|} \Bigg\{\left| v_j^{(k-1)} - \tilde{v}_j^{(k-1)}\right| + \frac{h_{jj}^2}{\left|D_{u_j}^{(k)}\right|} \left| a_j^{(k)}\{v_j^{(k-1)}\}^3 - \tilde{a}_j^{(k)}\{\tilde{v}_j^{(k-1)}\}^3\right| \Bigg\} \\ &\hspace{5.0cm} + 
     \Bigg\{ \frac{1}{\left|D_{u_j}^{(k)}\right|} \left|h_{ji}(h_{ji}\epsilon_{ji}+2h_{jj}\epsilon_{jj})v_{j}^{(k-1)}\right| \\ &\hspace{7.0cm} + \frac{1}{\{D_{u_j}^{(k)}\}^2} \left|h_{ji}^{}h_{jj}^2(2h_{jj}^2\epsilon_{ji}^{}+3h_{ji}^2\epsilon_{jj}^{})\{v_{j}^{(k-1)}\}^3\right| \Bigg\} \\ &\hspace{9.0cm} + \mathcal{O}(\epsilon^2+\epsilon^3+\dots)\Bigg]\tag*{since $\tilde{h}_{ii} = h_{ii} + \epsilon_{ii}$} \\
     \intertext{Combining both,}
     |v_i^{(k)} - \tilde{v}_i^{(k)}| &\leq  \Bigg\{\frac{1}{\left|D_{v_i}^{(k)}\right|}\mathcal{E}_{1_i} + \frac{\left| N_{\tilde{v}_i}^{(k)}\right|}{\left|D_{v_i}^{(k)}\right| \left|D_{\tilde{v}_i}^{(k)}\right|}\mathcal{E}_{2_i}\Bigg\} \tag*{where,} \\
     \mathcal{E}_{1_i} &= \frac{h_{ii}^2}{\left|D_{u_i}^{(k)}\right|} \left\{ \left| v_i^{(k-1)} - \tilde{v}_i^{(k-1)}\right| + \frac{h_{ii}^2}{\left| D_{u_i}^{(k)}\right|} \left|{a_i^{(k)} \{v_i^{(k-1)}\}^3} - \tilde{a}_i^{(k)} \{\tilde{v}_i^{(k-1)}\}^3\right| \right\}\\ &\hspace{5.0cm} + \frac{1}{\left|D_{u_i}^{(k)}\right|}\left\{\left| 2h_{ii} \epsilon_{ii}v_i^{(k-1)}\right| + \frac{1}{\left| D_{u_i}^{(k)}\right|}\left| 2h_{ii}^3 \epsilon_{ii}\tilde{a}_i^{(k)}\{v_i^{(k-1)}\}^3 \right|\right\} + \mathcal{O}(\epsilon^2+\epsilon^3+\dots)\\
     \mathcal{E}_{2_i} &= \mathlarger{\mathlarger{\sum}}_j \Bigg[ \frac{h_{ji}^2 \left|h_{jj}\right|}{\left|D_{u_j}^{(k)}\right|} \Bigg\{\left| v_j^{(k-1)} - \tilde{v}_j^{(k-1)}\right| + \frac{h_{jj}^2}{\left|D_{u_j}^{(k)}\right|} \left| a_j^{(k)}\{v_j^{(k-1)}\}^3 - \tilde{a}_j^{(k)}\{\tilde{v}_j^{(k-1)}\}^3\right| \Bigg\} \\ &\hspace{5.0cm} + 
     \frac{1}{\left|D_{u_j}^{(k)}\right|}\Bigg\{ \left|h_{ji}(h_{ji}\epsilon_{ji}+2h_{jj}\epsilon_{jj})v_{j}^{(k-1)}\right| \\ &\hspace{7.0cm} + \frac{1}{\left|D_{u_j}^{(k)}\right|} \left|h_{ji}^{}h_{jj}^2(2h_{jj}^2\epsilon_{ji}^{}+3h_{ji}^2\epsilon_{jj}^{})\{v_{j}^{(k-1)}\}^3\right| \Bigg\} \Bigg] \\ &\hspace{9.0cm} + \mathcal{O}(\epsilon^2+\epsilon^3+\dots)
     \intertext{Representing the linear terms by $\ell(\eps,\cdot)$, we get the desired result. }
 \end{align*}

\end{document}